\documentclass[aip,jcp,reprint]{revtex4-1}
\usepackage{graphicx,amsmath,verbatim,color}
\begin{document}

\title{Winding angles of long lattice walks}
\author{Yosi Hammer}
\email{hammeryosi@gmail.com}
\author{Yacov Kantor}
\affiliation{Raymond and Beverly Sackler School of Physics and
Astronomy, Tel Aviv University, Tel Aviv 69978, Israel}
\date{\today}

\begin{abstract}
We study the winding angles of random and self-avoiding walks on square and cubic lattices with number of steps $N$ ranging up to $10^7$. We show that the mean square winding angle $\langle\theta^2\rangle$ of random walks converges to the theoretical form when $N\rightarrow\infty$. For self-avoiding walks on the square lattice, we show that the ratio $\langle\theta^4\rangle/\langle\theta^2\rangle^2$ converges slowly to the Gaussian value 3. For self avoiding walks on the cubic lattice we find that the ratio $\langle\theta^4\rangle/\langle\theta^2\rangle^2$ exhibits non-monotonic dependence on $N$ and reaches a maximum of 3.73(1) for $N\approx10^4$. We show that to a good approximation, the square winding angle of a self-avoiding walk on the cubic lattice can be obtained from the summation of the square change in the winding angles of $\ln N$ independent segments of the walk, where the $i$-th segment contains $2^i$ steps. We find that  the square winding angle of the $i$-th segment increases approximately as $i^{0.5}$, which leads to an increase of the total square winding angle proportional to $(\ln N)^{1.5}$.
\end{abstract}

\maketitle

\section{Introduction} \label{sec:intro}

\begin{figure}
\includegraphics{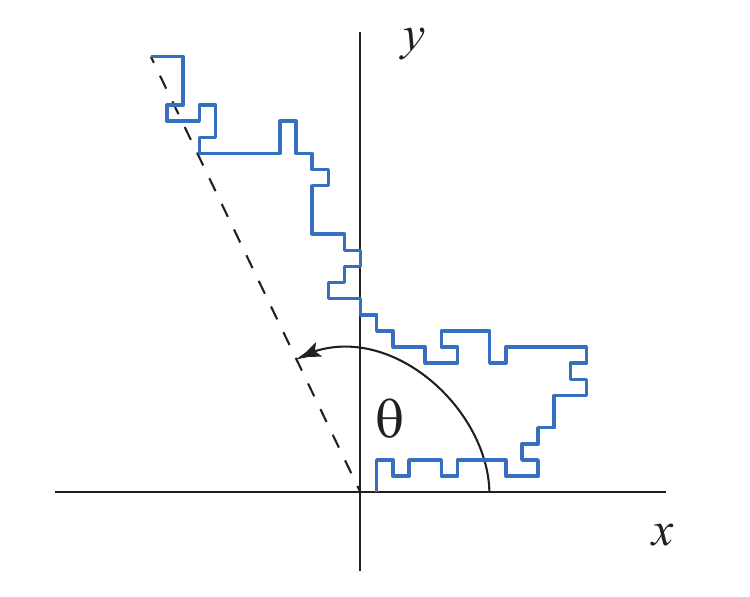}
\caption{Winding angle of a SAW on a square lattice.}
\label{fig:windAng}
\end{figure}

The winding properties of random walks (RWs) or self-avoiding walks (SAWs) around a point (in space dimension $d=2$) or a line (in $d=3$) have been studied extensively over the last sixty years. \cite{Spitzer1958,Edwards1967,Messulam1982,Fisher1984,Rudnick1987,Rudnick1988,Belisle1989,1991,Saleur1994,Drossel1996,
Samokhin1998,Grosberg2003,Walter2011,Walter2013,Belotserkovskii2014,Walter2014,Duplantier1988} 
The problem has implications in various fields of statistical physics, such as the conformations and dynamics of polymer chains \cite{Walter2011,Walter2013,Belotserkovskii2014,Walter2014} and flux lines in superconductors.\cite{Drossel1996} 
The first result by Spitzer\cite{Spitzer1958} showed that the probability distribution for the winding angle $\theta$ of a planar Brownian path around a point for large times $t$ is
\begin{equation} \label{eq:spitzer}
\lim_{t\rightarrow\infty} p\left(x=\frac{2\theta}{\ln t}\right) = \frac{1}{\pi}\frac{1}{1+x^2},
\end{equation} 
which is rather pathological since the averages $\langle|\theta|\rangle$ and $\langle\theta^2\rangle$ are both infinite due to the slow $x-$decay in Eq~\eqref{eq:spitzer}. The non-physical behavior originates from the fact that when a segment of the Brownian path approaches a \emph{point} center, it can wind around it an infinite amount of times. While this is true in an idealized system, in reality one expects that the polymer or Brownian particle will not be able to get infinitely close to the excluded center and a minimal distance will be imposed. For a lattice walk the cutoff distance is of the order of the lattice constant. When the minimal distance is incorporated into the model, the winding angle distribution in large $t$ becomes \cite{Rudnick1987}
\begin{equation} \label{eq:rudnick}
\lim_{t\rightarrow\infty} p\left(x=\frac{2\theta}{\ln t}\right) = \frac{\pi}{4\cosh^2(\pi x/2)}.
\end{equation}
In a lattice version of diffusion the diffusion time $t$ is proportional to the number of steps $N$ in the walk. (In our simulation $N$ will represent the number of \textit{sites} in the walk, but in the text we disregard this distinction.) The results in Eqs.~\eqref{eq:spitzer} and \eqref{eq:rudnick} were derived and verified by several methods, e.g., diffusion equation \cite{Spitzer1958,Edwards1967} and conformal mapping.\citep{Drossel1996}

For planar self-avoiding walks,\cite{Fisher1984,Rudnick1988,Duplantier1988}(Fig.~\ref{fig:windAng}) it was shown using conformal invariance \cite{Duplantier1988} that the winding angle follows a Gaussian distribution,
\begin{equation} \label{eq:Gaussian}
\lim_{N\rightarrow\infty}p\left(x=\frac{\theta}{2\sqrt{\ln N}}\right)=\frac{e^{-x^2}}{\sqrt{\pi}}.
\end{equation}

In three dimensions, the winding of a RW around an infinite line is practically identical with the two dimensional case since the steps in the plane perpendicular to the line are independent of the steps in the parallel direction. Therefore, the same distribution is expected for long walks. The problem of a SAW winding around a line, however, is more complicated. Rudnick and Hu \cite{Rudnick1988} considered a self-avoiding walk in $d=4-\epsilon$ and found that to first order in $\epsilon$, 
\begin{equation} \label{eq:epsilonEx}
p\propto \exp\left(-\frac{\theta^2\epsilon}{8\ln N}\right).
\end{equation}
Surprisingly, this result coincides with Eq.~\eqref{eq:Gaussian} for $\epsilon=2$. However, no exact result is known for the distribution of a self-avoiding walk around a rod. Moreover, in recent simulations by Walter \textit{et al.},\cite{Walter2011} it was found that $p(\theta)$ decreases slower than a Gaussian function, at odds with the first order $\epsilon$-expansion results.

Due to the slow approach of the distribution to the asymptotic form (such as $\langle\theta^2\rangle\sim\ln N$ for planar self-avoiding walks), in order to verify the results in Eqs.~\eqref{eq:rudnick}, \eqref{eq:Gaussian} and  \eqref{eq:epsilonEx} in simulations, one has to use very long random and SAWs. This can be challenging since traditionally, in order to measure the winding angle of an $N$-step walk, one has to trace all the sites visited by the walk, which takes time of $O(N)$. Note that the winding angle $\theta$ is the \emph{total accumulated angle} of the steps along the walk. (It is \emph{not} defined modulus $2\pi$.) Moreover, the generation of a large ensemble of long SAWs is difficult on its own, due to the need to check for intersections of the walk with itself. In this work we improve upon known measurements of the winding angle of RWs and SAWs by using a new implementation of the pivot algorithm that was introduced by Clisby in recent years.\cite{Clisby2010,Clisby2010b} The implementation allows us to generate a large ensemble of RWs and SAWs with up to $\sim10^7$ steps and compute their winding angle without writing down the entire walk, so that a measurement of the winding angle of a walk with $N$ sites is done in time of $O(\ln N)$.

\section{Efficient measurement of global properties of polymers} \label{sec:algo}
As mentioned above, Monte Carlo simulations face a challenge to generate large ensembles of SAWs. The pivot algorithm \cite{Lal1969,Madras1988,Kennedy2002,Clisby2010,Clisby2010b} is a dynamic method which generates SAWs with fixed $N$ and free end-points. At each time step a random site along the walk is used as a pivot point for a random symmetry action on the lattice (e.g., rotation or reflection) to be applied to the part of the walk subsequent to the pivot point. The resulting walk is accepted if it is self-avoiding; otherwise, it is rejected and the old walk is sampled again. The pivot algorithm is most efficient when studying large scale properties of the polymers.\cite{Madras1988} In the past, the bottleneck of the algorithm was the self-avoidance tests, which required $O(N^x)$ operations ($x\cong1/2$). 

\begin{figure}
\includegraphics{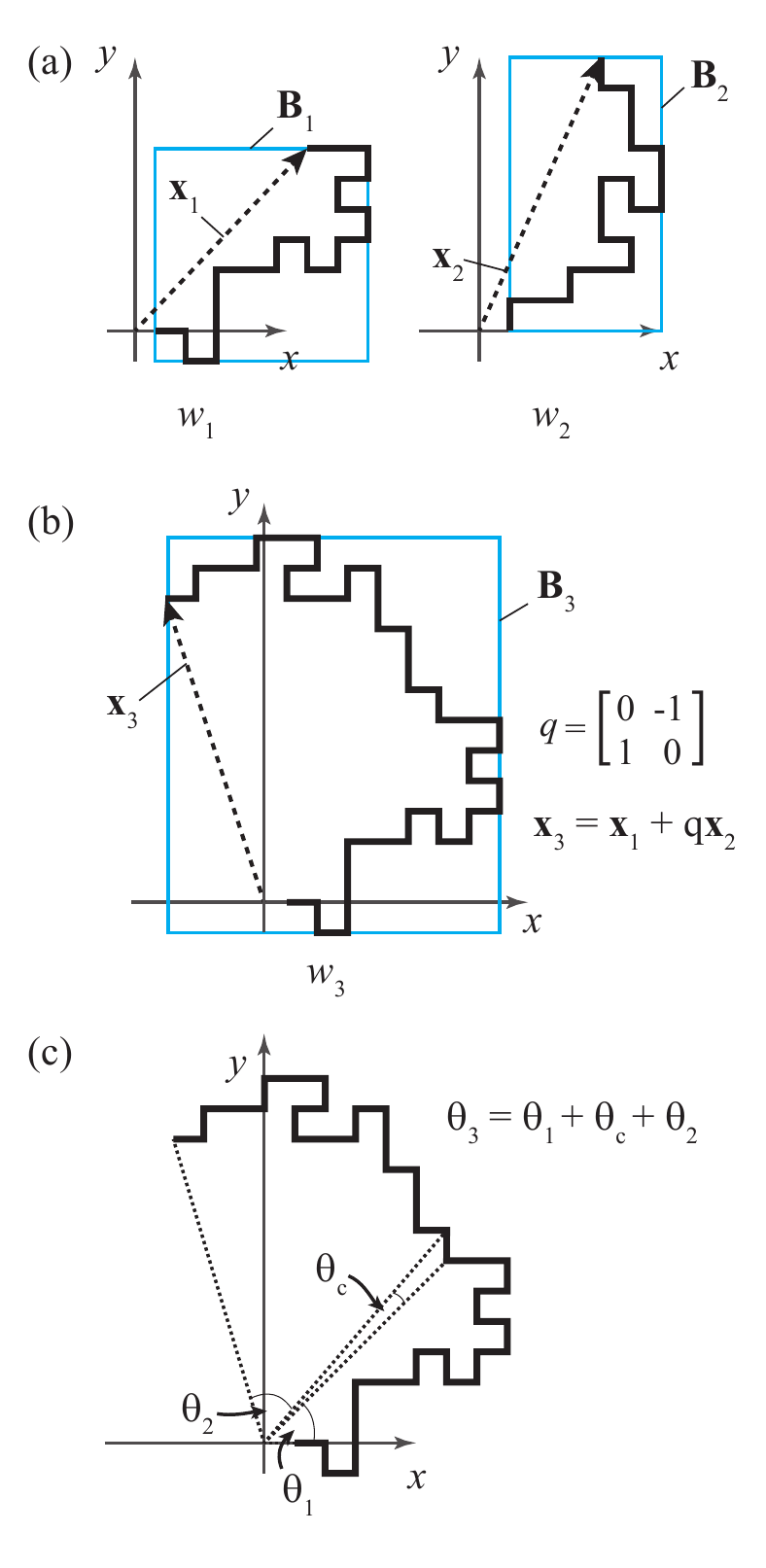}
\caption{(Color online) (a) Two SAWs, $w_1$ and $w_2$, drawn on a square lattice. By convention the walks start at $(1,0)$. The end points of the walks are denoted $\mathbf{x}_1$ and $\mathbf{x}_2$, and are indicated by the dashed lines. The bounding boxes of the walks are marked by the cyan rectangles. (b) The walk $w_3$ is obtained by using a symmetry operation $q$ on $w_2$ and then concatenating it with $w_1$. The end point $\mathbf{x}_3$ can be obtained from $\mathbf{x}_1$, $\mathbf{x}_2$ and $q$ without knowing the position of all the sites along the walks $w_1$ and $w_2$. (c) The winding angle $\theta_3$ of $w_3$ can also be derived from the global properties of its sub-walks. The angles $\theta_1$, $\theta_2$ and $\theta_c$ can be computed from the knowledge of $\mathbf{x}_1$, $\mathbf{x}_2$ and $q$, without tracing all the steps along the walk.}
\label{fig:recursiveStructure}
\end{figure}

Clisby managed to drastically improve the efficiency of the pivot algorithm\cite{Clisby2010} so that a pivot attempt is done in a time not exceeding $O(\ln N)$. He accomplished this by storing the walks in a new data structure in which a walk is represented as the concatenation of sub-walks of smaller sizes. A global property of the walk can be deduced from the properties of the sub-walks it is constructed from. For example, the end-point of the walk can be found by using the end-points of each of the sub-walks, along with the symmetry operation used during the concatenation. This is illustrated in Fig.~\ref{fig:recursiveStructure}. In Fig.~\ref{fig:recursiveStructure}(a) two SAWs on a square lattice, $w_1$ and $w_2$, are drawn. The SAW in Fig.~\ref{fig:recursiveStructure}(b) is obtained by applying the symmetry operation $q$ (a 90 degree counter-clockwise rotation) to $w_2$ and then concatenating it with $w_1$. Note that we use the convention that all walks start from $(1,0)$. The end point $\mathbf{x}_i$ of $w_i$ is marked by a dashed line in Figs.~\ref{fig:recursiveStructure}(a) and \ref{fig:recursiveStructure}(b). It is shown that $\mathbf{x}_3$ can be derived from $\mathbf{x}_1$, $\mathbf{x}_2$ and $q$, without knowing the positions of all the sites in the walk. 

The data structure used to represent the walks in the simulation is a binary tree where each node contains the global properties of the walk which corresponds to sub-tree that contains all the nodes below it in the tree. These properties include the end point of the walk, the symmetry operation used to concatenate its sub-walks and a bounding box. The bounding box is a convex shape that completely contains the walk (see Fig.~\ref{fig:recursiveStructure}(a),(b)). Clisby showed that pivot operations can be done by applying transformations to change the structure of the tree and the symmetry operations in the nodes. Moreover, self intersection tests can be done by recursively checking for intersections between the bounding boxes of right and left children in the tree. These procedures are explained in detail in Ref.~\onlinecite{Clisby2010}. 

In our simulation, we use the fact that the winding angle of a random or SAW is also a global property that can be deduced from the sub-walks that form it. Consider a random or SAW $w$ which is represented by Clisby's binary tree. The following recursive function can be used to compute its winding angle: 
\subsubsection*{\emph{Compute $\theta(w)$}}
\begin{figure}
\includegraphics{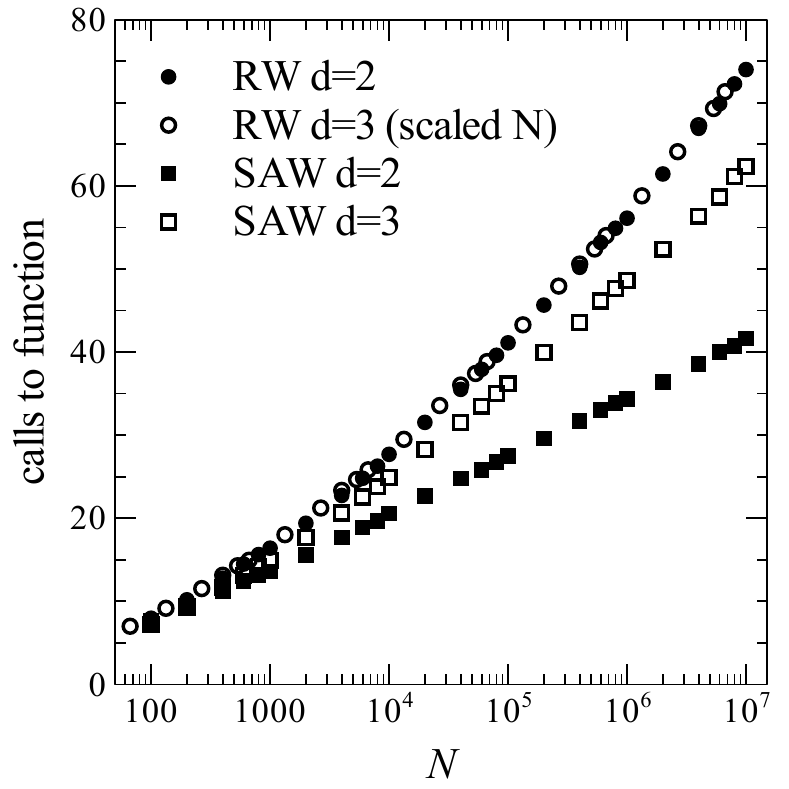}
\caption{The average number of times the recursive function \emph{Compute $\theta(w)$} that computes the winding angle of an $N$-step walk $w$ is called as a function of the number of sites $N$ of the walk. The circles denote RWs and the squares denote SAWs. Open symbols correspond to dimension $d=3$ and full symbols to $d=2$. Note that for RWs in $d=3$, we re-scaled $N$ by a factor of $2/3$, the average fraction of steps taken in a direction perpendicular to the $z$ axis.}
\label{fig:numOfCalls} 
\end{figure}

\begin{enumerate}
\item Check whether the bounding box of $w$ intersects with the line (in dimension $d=3$) or point (in $d=2$) $x=y=0$. 
\item If not, there is no possibility that the walk has encircled the origin and the winding angle can be computed immediately from the position of the first site and the end point of the walk. The function will then return this angle and terminate. Note that the necessary information is found in the node at the root of the tree and there is no need to trace the sites along the walk. 
\item If the bounding box does intersect with the line/point $x=y=0$:
\begin{enumerate}
\item Call the function again to calculate $\theta_1=$ \emph{compute $\theta(w_l)$}, where $w_l$ is the sub-walk of $w$ that corresponds to the left sub-tree of the SAW tree which represents $w$.
\item Call the function again to calculate $\theta_2=$ \emph{compute $\theta(w_r)$}, where $w_r$ is defined similarly to $w_l$. Note that when computing $\theta_2$ we need to take into account the fact that $w_r$ is acted upon by a symmetry operation $q$ and then shifted to the end point of $w_l$, $\mathbf{x}_l$, before it is concatenated with $w_l$.
\item Calculate the angle $\theta_c$, between the last site of $w_l$ and the first site of $w_r$. This angle will depend only on $\mathbf{x}_l$ and $q$.
\end{enumerate} 
\item return $\theta_1+\theta_c+\theta_2$.
\end{enumerate}
The operation of \emph{Compute $\theta(w)$} is illustrated in Fig.~\ref{fig:recursiveStructure}(c). In order to compute the winding angle of $w_3$, the function will first check whether its bounding box intersects with the origin. Since the bounding box of $w_3$ \emph{does} intersect with the origin [see Fig.~\ref{fig:recursiveStructure}(b)]. The function will continue to compute $\theta_1$ and $\theta_2$ recursively and the angle $\theta_c$ shown in the figure and return their sum. Since the lattice walks tend to move away from the origin, when the function \emph{Compute $\theta(w)$} is called to compute the winding angle of sections of the walk that are far from its beginning, it will usually find that their bounding boxes do not intersect with the point/line $x=y=0$, and return their winding angle in a single step. In fact, we find that the number of times that \emph{Compute $\theta(w)$} is called is very small even when the number of sites in $w$ is large. This is illustrated in Fig.~\ref{fig:numOfCalls}. For example, in order to compute the winding angle of a SAW in $d=3$ with $N=10^7$, the recursive function is called only 62 times on average.

We studied the winding angles of RWs and SAWs of sizes ranging from $N=100$ to $N=10^7$ sites on square and cubic lattices. For each size, we started from an initial configuration, performed a sequence of pivot attempts and computed their winding angles after each attempt. The initial configuration was selected in the same way as in Ref.~\cite{Hammer2015}. If, as a result of a potential pivot move the walk intersects with the point/line $x=y=0$, or, for a SAW, intersects with itself, the pivot move is rejected and the same configuration is sampled again. An important detail in the implementation of the pivot algorithm is the distribution from which the pivot point along the walk is selected. Usually, the pivot point is selected uniformly from the sites along the walk. \cite{Madras1988} However, when computing the winding angle of the walks, we find that the correlation between successive measurements is significantly reduced when we use a distribution that favors sites closer to the starting point of the walk. We used the distribution that was defined in Ref.~\onlinecite{Clisby2016}.

\section{Winding angle of RWs} \label{sec:RWs}
In order to test the theoretical prediction for the winding angle of RWs we can compute the average square winding angle and check its dependence on $N$. From Eq.~\eqref{eq:rudnick},
\begin{equation}
\langle x^2 \rangle = \int x^2p(x)dx = \frac{1}{3},
\end{equation}
i.e., $\langle \theta^2 \rangle = (\ln t)^2/12$.
For a RW with large $N$, the winding angle is expected to agree with Eq.~\eqref{eq:rudnick} where the diffusion time $t$ is proportional to the number of steps in the walk, i.e., $t = c_0N$. We therefore expect that in the large $N$ limit, 
\begin{equation}
\label{eq:RWlinear}
\sqrt{\langle\theta^2}\rangle = \frac{1}{\sqrt{12}}(\ln c_0 + \ln N) = A + \frac{1}{\sqrt{12}}\ln N.
\end{equation}
The additive constant $A$ in Eq.~\eqref{eq:RWlinear} depends on local properties of the system like the shape and size of the excluded area near the origin where the RW is not allowed to visit.\cite{Drossel1996} However, in the large $N$ limit, the root mean square winding angle should be proportional to $\ln N$ with a prefactor of $1/\sqrt{12}$, independent of any local properties. 

In Ref.~\onlinecite{Rudnick1987}, $\sqrt{\langle\theta^2}\rangle$ was studied as a function of $N$ for RWs on a square and cubic lattices with $N$ ranging up to $10^3$. Some non-negligible deviations from the theory were observed, and it was stipulated that the main reason for these was finite size corrections. Here we use walks with $N=10^2-10^7$ to study the effect of finite $N$ and see if $\sqrt{\langle\theta^2}\rangle$ converges to the predicted form when $N$ increases.

\begin{figure}
\includegraphics{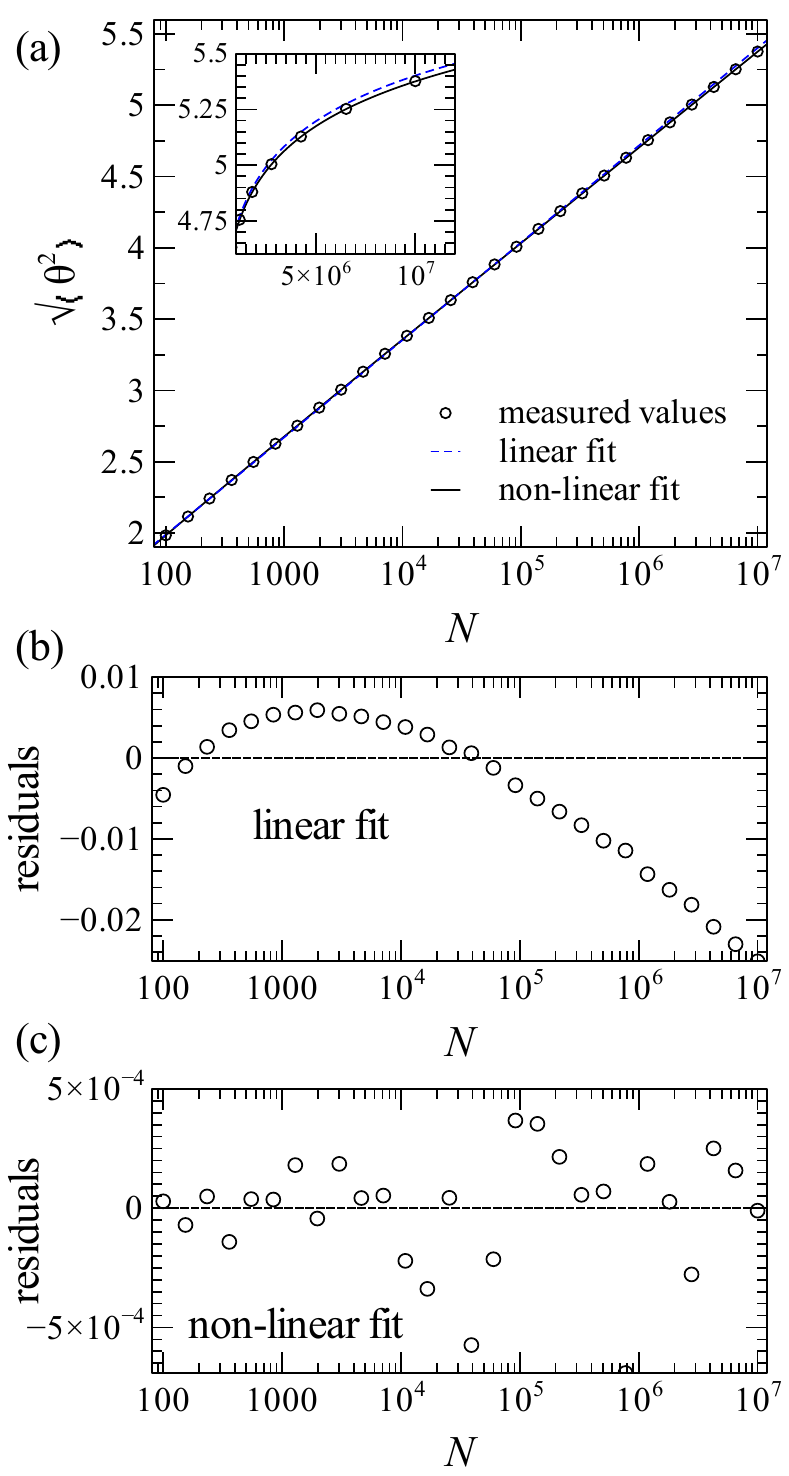}
\caption{(Color online) (a) Root mean square winding angle of a RW on a square lattice, shown as a function of the number of sites in the walk $N$ in semi-logarithmic scale. The region with $N=10^6-10^7$ is shown in the inset in linear scale. The dashed line denotes a linear fit and the solid line denotes non-linear fits [Eqs.~ \eqref{eq:RW2dNonLinFit} and \eqref{eq:RW2dNonLinFit2}] that take into account finite size corrections. (b) Residuals of the linear fit. (c) Residuals of the non-linear fits. Note that the curves from Eqs.~ \eqref{eq:RW2dNonLinFit} and \eqref{eq:RW2dNonLinFit2} are indistinguishable in this range of values for $N$.}
\label{fig:2dRW}
\end{figure}
In Fig.~\ref{fig:2dRW}(a), $\sqrt{\langle\theta^2}\rangle$ is plotted against $N$ in semi-logarithmic scale. The dependence is very close to linear. In order to take into account the finite size corrections, we performed two non-linear fits. The first was to the form
\begin{equation} \label{eq:RW2dNonLinFit}
\sqrt{\langle\theta^2\rangle} = A + B\ln N + C(\ln N)^{-\Delta},
\end{equation}
and in the second we used integer powers of $1/\ln N$, i.e., 
\begin{equation} \label{eq:RW2dNonLinFit2}
	\sqrt{\langle\theta^2\rangle} = c_0 + c_1\ln N + c_2(\ln N)^{-1}+c_3(\ln N)^{-2}.
\end{equation}
The curves produced by the two forms are practically indistinguishable for the values of $N$ in our simulation. For a RW on a square lattice, this curve is denoted by the solid line in Fig.~\ref{fig:2dRW}(a), where it is shown that it is in a slightly better agreement with the data than the linear fit denoted by the dashed line. In Figs.~\ref{fig:2dRW}(b) and \ref{fig:2dRW}(c) we present the residuals of the linear and non-linear fits. Clearly, the linear form suffers from a systematic disagreement with the data, while the residuals of the non-linear form seem to scatter randomly around zero. Thus, both Eq.~\eqref{eq:RW2dNonLinFit} and Eq.~\eqref{eq:RW2dNonLinFit2} agree with the results of our simulations. A similar behavior was observed for RWs on the cubic lattice. The fitting parameters acquired from the non-linear fits are presented in Table \ref{tab:2dRWfitPar}. Note that the parameter $B$ from the fit to Eq.~\eqref{eq:RW2dNonLinFit} is in agreement with the theoretical value of $1/\sqrt{12}\approx0.2887$, both in dimensions $d=2$ and $d=3$, while $c_1$ from the fit to Eq.~\eqref{eq:RW2dNonLinFit2} slightly deviates from the theory. Possibly, Eq.~\eqref{eq:RW2dNonLinFit} captures the leading order finite size corrections to $\sqrt{\langle\theta^2\rangle}$ more accurately.

\setlength{\tabcolsep}{12pt}
\begin{table}
\begin{tabular}{c  c  c}
 & $d=2$ & $d=3$ \\ \hline \hline
$A$ & $0.78 \pm 0.03$ & $0.69 \pm 0.03$  \\
$B$ & $0.2888 \pm 0.0005$ & $0.288 \pm 0.001$ \\
$C$ & $-0.321 \pm 0.017$ & $-0.357 \pm 0.012$ \\
$\Delta$ & $0.62 \pm 0.12$ & $0.54 \pm 0.11$ \\
$c_0$ & $0.74\pm0.006$ & $0.63\pm0.005$\\ 
$c_1$ & $0.2894\pm0.0003$ & $0.2892\pm0.0002$\\
$c_2$ & $-0.48\pm0.05$ & $0.56\pm0.04$\\
$c_3$ & $0.33\pm0.11$ & $0.45\pm0.09$ \\
\end{tabular}
\caption{Fitting parameters from Eqs.~\eqref{eq:RW2dNonLinFit} and \eqref{eq:RW2dNonLinFit2} for the root mean square winding angle of RWs with $N$ sites on square ($d=2$) and cubic ($d=3$) lattices.}
\label{tab:2dRWfitPar}
\end{table}

\section{SAWs on the square lattice} \label{sec:SAW2d}
For long SAWs on a square lattice, the winding angle was shown to have a Gaussian distribution [Eq.~\eqref{eq:Gaussian}].\cite{Duplantier1988} Even before this analytical result, the winding angle was studied numerically. Fisher \textit{et al.}\cite{Fisher1984} used exact enumeration of short (up to $N=21$) walks and Monte Carlo simulations of SAWs with $N\leq170$ to measure the winding angle distribution. They found that, to a good approximation, 
\begin{equation} \label{eq:SAWlinear}
\langle\theta^2\rangle \propto \ln N,
\end{equation}
and
\begin{equation} \label{eq:fisherResult}
2.9<\frac{\langle\theta^4\rangle}{\langle\theta^2\rangle^2}<3.2,
\end{equation}
which is close to the Gaussian value 3. Due to the limited computer resources at the time and the lack of an efficient algorithm to compute the winding angle of the walks, it was not possible to measure the winding angles longer SAWs and observe the convergence to the Gaussian form. 

\begin{figure}
\includegraphics{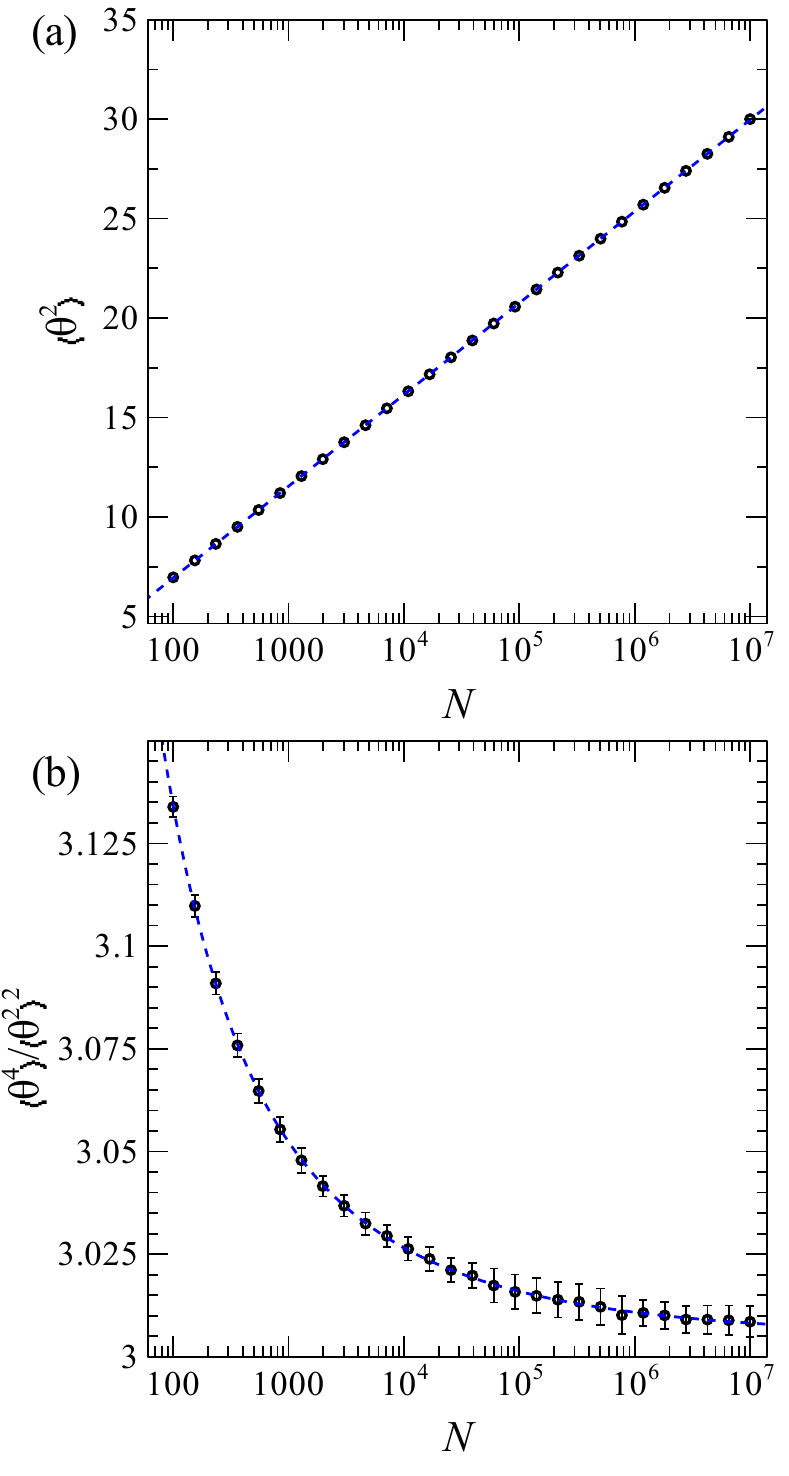}
\caption{(Color online) (a) The mean square winding angle of a SAW with $N$ sites on a square lattice. The dashed line is a linear fit. (b) The ratio $\langle\theta^4\rangle/\langle\theta^2\rangle^2$ approaches the Gaussian value 3 as $N$ increases. The dashed line denotes a non-linear fit according to Eq.~\eqref{eq:2dSAWNonLinFit}}
\label{fig:SAW2d}
\end{figure}
In this work, we studied the winding angles of SAWs on the square lattice with $N$ ranging from $10^2$ to $10^7$. The mean square winding angle is shown as a function of $N$ in semi-logarithmic scale in Fig.~\ref{fig:SAW2d}(a). The dashed line denotes a linear fit which results in 
\begin{equation} \label{eq:2dSAWangSqrFit}
\langle\theta^2\rangle = (-2.268\pm0.002) + (2\pm0.002)\ln N.
\end{equation}
The slope is in excellent agreement with the theoretical value [Eq.~\eqref{eq:Gaussian}]. Note that the linear relation holds quite well even for relatively short walks ($N\sim10^2$). To observe finite size corrections to the Gaussian distribution of $\theta$, we measured the ratio $\langle\theta^4\rangle/\langle\theta^2\rangle^2$. The results are shown in Fig.~\ref{fig:SAW2d}(b). Note that (a) even for $N=10^2$, $\langle\theta^4\rangle/\langle\theta^2\rangle^2\approx3.13$, not far from the Gaussian value, which is consistent with the results for the linear dependence of $\langle\theta^2\rangle^2$ on $\log N$. (b) The small but noticeable difference from the Gaussian form converge slowly to zero. Even for walks with $N=10^7$ we observe a non-negligible deviation from 3. To study the finite size corrections we fitted the data to the form
\begin{equation} \label{eq:2dSAWNonLinFit}
\langle\theta^4\rangle/\langle\theta^2\rangle^2=A+B(\ln N)^{-1}+C(\ln N)^{-2}.
\end{equation}
[This form was in slightly better agreement with the data than a function with non-integer powers like in Eq. \eqref{eq:RW2dNonLinFit}.] The result is denoted by the dashed line in Fig.~\ref{fig:SAW2d}(b). We find $A = 3.0097 \pm 0.0013$, $B = -0.262 \pm 0.020$, and $C= 3.85 \pm 0.07$. Note that $A$ is very close to the Gaussian value. The small difference is most likely a result of corrections in the form of higher powers of $1/\ln N$ that were not taken into account in Eq.~\eqref{eq:2dSAWNonLinFit}. 

\section{SAWs on the cubic lattice} \label{sec:SAW3d}
Despite the long standing interest in this problem, the exact distribution $p_N(\theta)$ of the winding angle of a SAW with $N$ sites around a rod in dimension $d=3$ is unknown. The only analytical result that we know of was obtained by Rudnic and Hu,\cite{Rudnick1988} where they used renormalization group methods to show that in $d=4-\epsilon$, to first order in $\epsilon$, $p_N(\theta)$ follows the Gaussian distribution given in Eq.~\eqref{eq:epsilonEx}. The authors also reported a Monte Carlo simulation of SAWs with up to 910 steps, where they were only able to study the pre-asymptotic regime where RW behavior was observed. More recently, Walter \textit{et al.}\cite{Walter2011} utilized the improvement in computer power to study the winding angle distribution of SAWs on the cubic lattice with $N\le25000$. They showed that $p_N(\theta)$ does not converge to the Gaussian form and found that $\langle\theta^2\rangle\propto(\ln N)^{2\alpha}$ where $\alpha=0.75(1)$. They also showed that $\langle\theta^4\rangle/\langle\theta^2\rangle^2$ converges to 3.74(5), which differs significantly from the first order $\epsilon$-expansion prediction.
Here we extend the study to walks with $N$ up to $10^7$ to see whether the behavior that was observed in Ref.~\onlinecite{Walter2011} persists as $N\rightarrow\infty$, or a pre-asymptotic regime has been observed.
\begin{figure}
\includegraphics{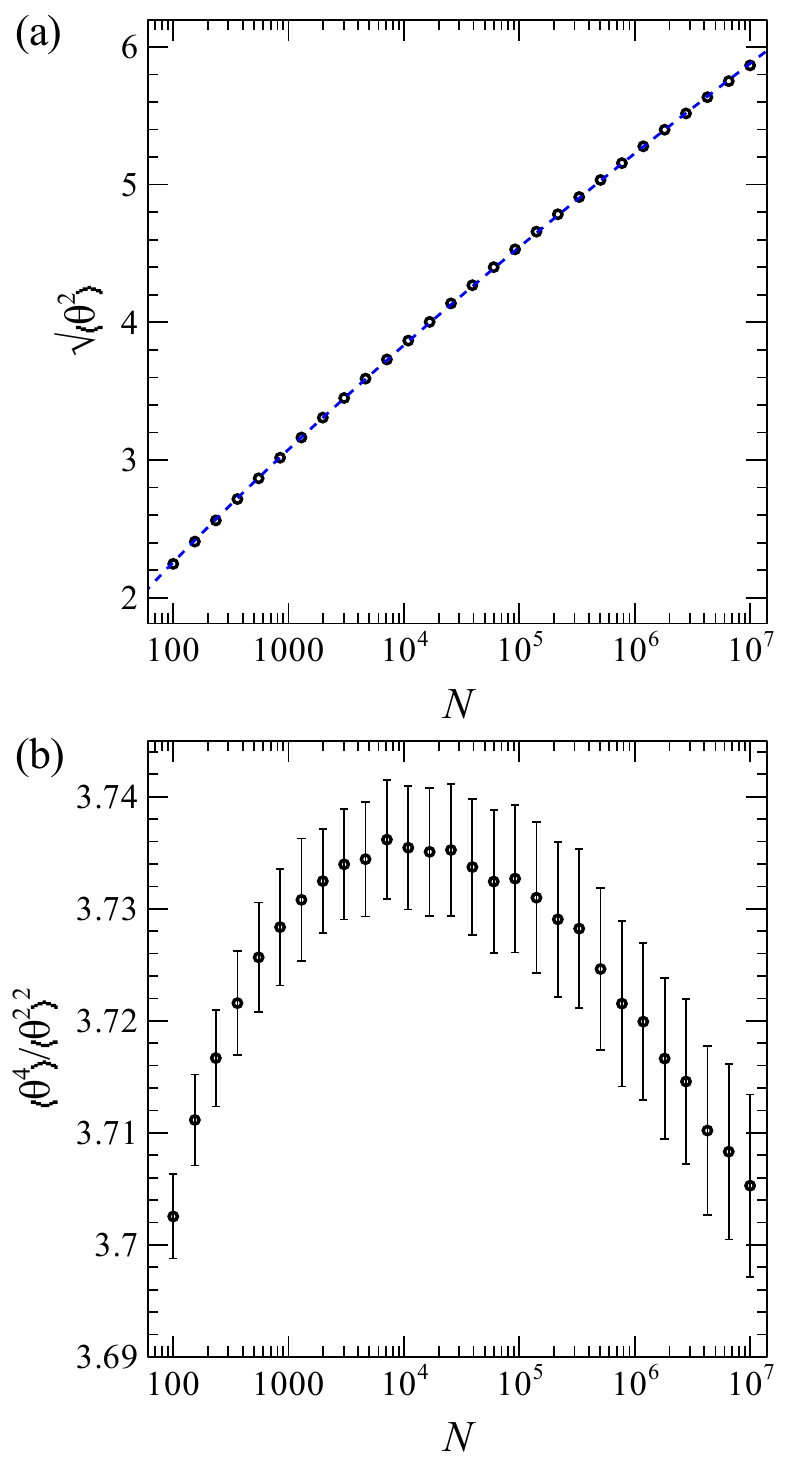}
\caption{(Color online) (a) The root mean square winding angle of a SAW with $N$ sites on a cubic lattice. The dashed line is a power law fit. (b) The ratio $\langle\theta^4\rangle/\langle\theta^2\rangle$.}
\label{fig:SAW3d}
\end{figure}

Our results for the root mean square winding angle of a SAW with $N$ sites on a cubic lattice are depicted in Fig.~\ref{fig:SAW3d}(a). The dashed line denotes a power law fit to the form
\begin{equation} \label{eq:SAW3dPower}
\sqrt{\langle\theta^2\rangle} = A\cdot (\ln N)^{\alpha},
\end{equation}
that resulted in $A=0.703\pm0.004$ and $\alpha=0.764\pm0.003$. Note that $\alpha$ extracted from our data is in agreement with the result in Ref.~\onlinecite{Walter2011}. The ratio $\langle\theta^4\rangle/\langle\theta^2\rangle^2$ from our simulation is shown in Fig.~\ref{fig:SAW3d}(b). Note the \textit{non-monotonic} dependence of $\langle\theta^4\rangle/\langle\theta^2\rangle^2$ on the number of sites in the walk. We find that it reaches a maximum value of $3.73\pm0.01$ for $N\approx10^4$ (in agreement with the result in Ref.~\onlinecite{Walter2011}) and then decreases as the walks increase in size. For $N=10^7$ we find $\langle\theta^4\rangle/\langle\theta^2\rangle=3.705\pm0.008$, still far from the Gaussian value 3 predicted from first order $\epsilon$-expansion.\cite{Rudnick1988} The non-monotonic behavior is quite surprising, since we do not have a reason to expect that $\langle\theta^4\rangle/\langle\theta^2\rangle^2$ will reach its maximum value around $N=10^4$. 
This behavior might indicate a cross-over from a non-asymptotic regime at finite $N$ to the asymptotic regime at infinite $N$.
\begin{figure}
	\includegraphics{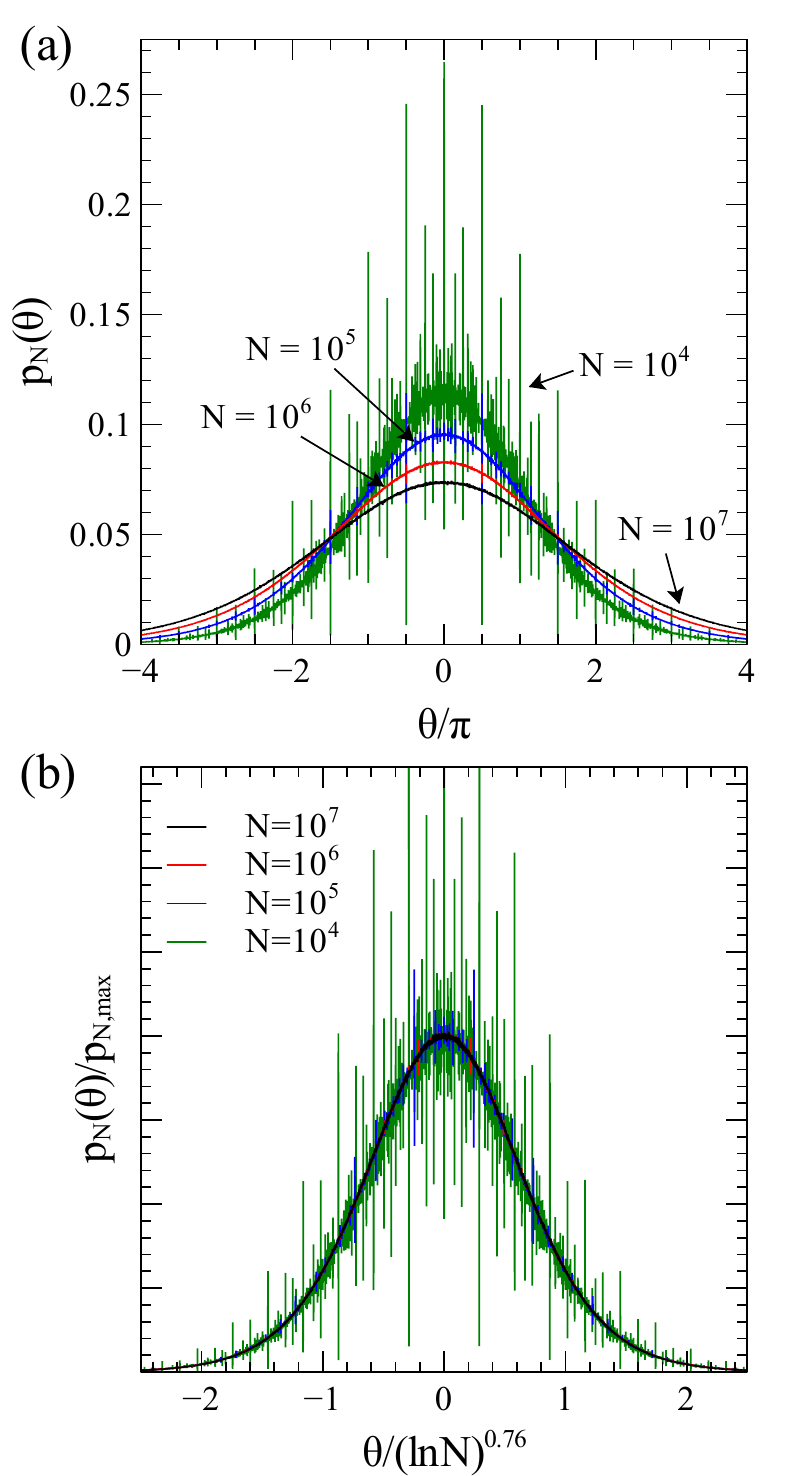}
	\caption{(Color online) (a) Winding angle distribution of a SAW with $N$ steps on a cubic lattice. (b) The distribution divided by the maximal value, as a function of scaled angle.}
	\label{fig:windAngDist}
\end{figure}
One possible reason for such a cross-over can be the influence of the lattice on which the walks are created. In Fig.~\ref{fig:windAngDist} we present the winding angle distribution of SAWs on a cubic lattice for $N$ ranging from $10^4$ to $10^7$. For $N=10^4$ the lattice structure is evident in the distribution $p_N(\theta)$, where narrow peaks are observed at specific angles corresponding to lattice sites close to the $z$ axis. This effect is diminished significantly when $N$ increases, and the peaks in the distribution are not observable in our simulation for $N=10^7$. In Fig.~\ref{fig:windAngDist}(b) we show that $p_N(\theta)$ for different $N$ collapse to a single curve when $\theta$ is scaled by $(\ln N)^{0.76}$, in agreement with the results in Ref.~\onlinecite{Walter2011} and with the behavior shown in Fig.~\ref{fig:SAW3d}. Note that the effects of the lattice structure were not evident in Figs. 8 and 9 in Ref.~\onlinecite{Walter2011} due to the coarse binning in those graphs (The bin width in the histograms was 0.5 radians, compared to $\pi/1000$ in our graphs.).

\section{The Gaussian argument}
\label{sec:gussian}
The Gaussian form of the winding angle distribution given in Eq.~\eqref{eq:Gaussian} can be explained by the following simple argument:\cite{Drossel1996,Fisher1984} Starting from the first step, the SAW can be divided into segments of lengths $1,2,4,...,2^m\approx N/2$. The $i$-th segment has $2^i$ steps, and starts after approximately $2^i$ steps of the walk. Thus both its linear size and its distance from the origin are typically of the order of $2^{i\nu}$, where $\nu=3/4$ for SAWs in $d=2$.\cite{Gennes1979} The winding angle of each segment is then expected to be of order one. The total winding angle of the walk is approximately the sum of the changes in the winding angles of these individual segments, i.e., 
\begin{equation} \label{eq:selfSimilarity}
	\theta = \sum_i \Delta\theta_i.
\end{equation}
The SAW is a self-similar object, and it is expected that the properties of the smaller segments are identical to those of the larger segment when they are scaled down to the same size. Thus, the angles $\Delta\theta_i$ have identical distributions. Under the assumption that they are independent, and with a finite variance, the central limit theorem states that in the limit of large $m$, $\theta$ will have a Gaussian distribution with a variance proportional to $m\propto\ln N$. 

In Ref.~\onlinecite{Drossel1996} it was mentioned that this argument fails when it is applied to RWs since a RW is allowed to return to the vicinity of the excluded center, while a SAW in $d=2$ cannot return to the origin without self-intersection. In fact in $d=2$ the size of the effective excluded area is of the order of the segment size, and upon rescaling those properties remain unchanged. For a RW, the size of the excluded center has the size of a lattice cell for any segment size. We suspect that this is also the reason why the Gaussian argument fails for SAWs in $d=3$.

\begin{figure}
	\includegraphics{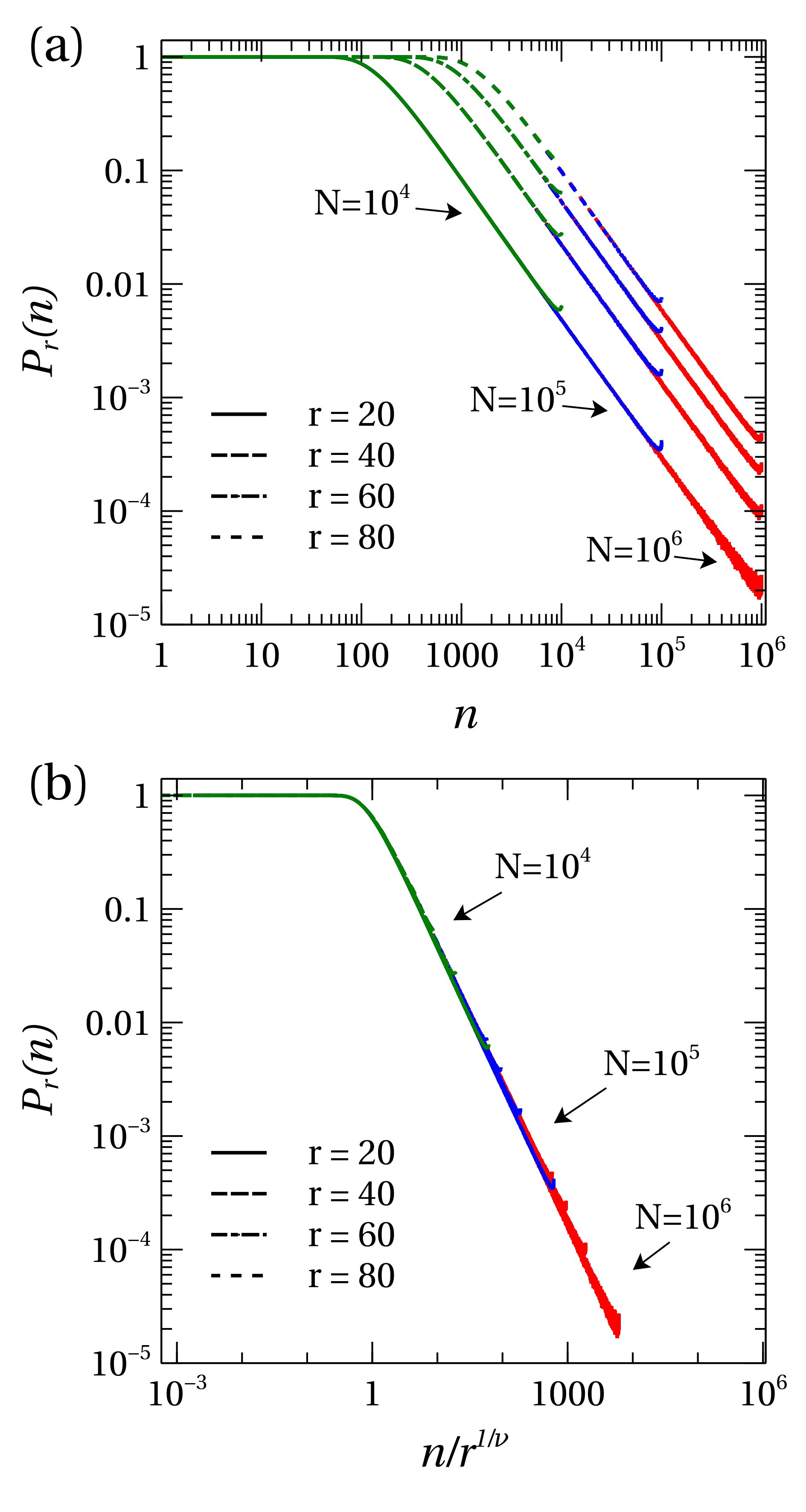}
	\caption{(Color online) Probability of the $n$ site in a SAW with $N$ steps on the cubic lattice to be inside a cylinder of radius $r$ around the $z$ axis.}
	\label{fig:cylProb}
\end{figure}

In order to understand the behavior of SAW in $d=3$ we studied their return to an infinite cylinder of radius $r$ centered on the $z$ axis. (The SAW tree allows very fast intersection tests between the walk and a the cylinder, that take time no longer than $O(\ln N)$).\cite{Hammer2015} In Fig.~\ref{fig:cylProb} we present the probability $P_r(n)$ that the site $n$ in a SAW of $N$ steps will be inside a cylinder of radius $r$. Note that in this simulation these cylinders are not excluded regions like the infinite line $x=y=0$ and we use them simply to check if the walk is wandering close to the excluded center. In Fig.~\ref{fig:cylProb}(a) we depict on a logarithmic scale $P_r(n)$ as a function of $n$ for several sizes $r$ and several total lengths $N$ of the SAWs. We note that for fixed $r$ the data points of larger $N$s nicely continue the trend of the smaller $N$s. This is true apart from a small increase in $P_r(n)$ near the end of the walk (as $n\sim N$). This increase implies that the probability for the site $n$ to be inside the cylinder is reduced by the presence of a \emph{finite} part of the walk subsequent to $n$, and when this part is not present (near the end of the walk) it is easier for the walk to return to the cylinder. All graphs have $P_r(n)\approx1$ until the size of the size of the segment of the walk $an^{\nu}\sim r$, and then decay with a slope of about -1.2. This decay is slow but converging in the sense that in the asymptotic limit (when $N\rightarrow\infty$) only a finite number of sites will be in the cylinder. In Fig.~\ref{fig:cylProb}(b) we show that these curves collapse when $P_r$ is plotted against the scaled position $n/r^{1/\nu}$. This collapse demonstrates that when a long walk is scaled down, the statistics of the walk in a cylinder around the $z$ axis correspond to those of a smaller cylinder, with reduced radius. The rescaling in Fig.~\ref{fig:cylProb}(b) does not change the size of the excluded line, but the collapse of varying cylinder radius to the same curve indicate that if we take a segment of a SAW and scale it down, it does not correspond to an earlier (smaller) segment but correspond to the behavior of a SAW with a smaller excluded center. Thus, its winding angle distribution will not be identical to that of a preceding segment and will, probably, have a larger variation $\Delta\theta$.

\begin{figure}
	\includegraphics{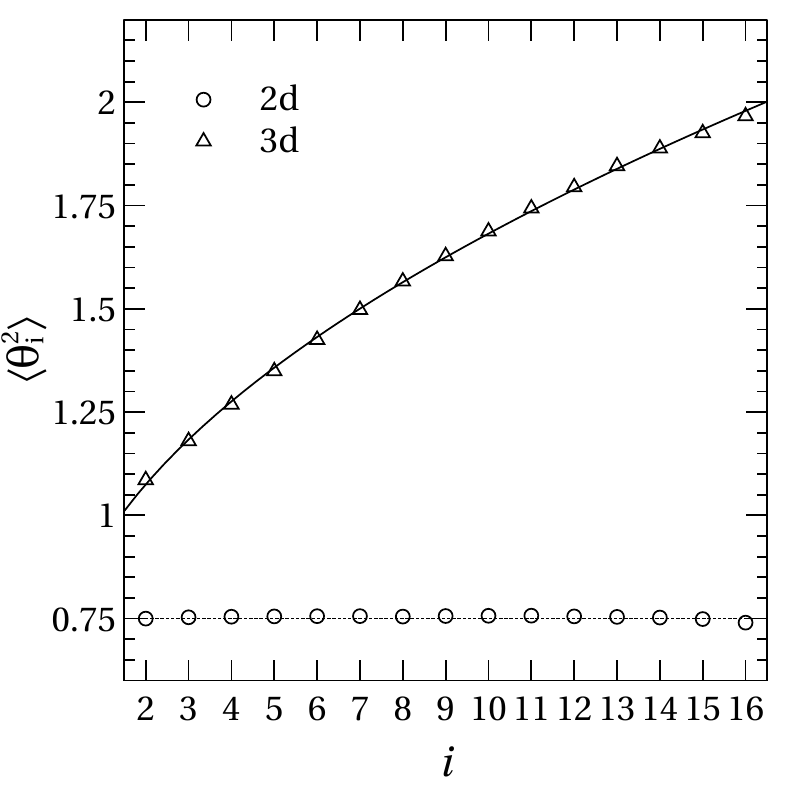}
	\caption{The change in the winding angle of of a SAWs on square ($d=2$) and cubic ($d=3$) lattices for segments of the walk of sizes $64*2^i$, starting from the beginning of the walk. The first segment ($i=1$) was omitted from the graph. For SAWs in $d=2$, $\langle\Delta\theta_i^2\rangle$ is approximately 0.75 for all segments (dashed line), while for SAWs in $d=3$, $\langle\Delta\theta_i^2\rangle$ increases with $i$. The solid line represents a power law increase of $\langle\Delta\theta_i^2\rangle\sim i^{0.52}$.}
	\label{fig:gaussian}
\end{figure}

By rotating the SAW tree we can divide the walk into segments of differing sizes in the simulation. For SAWs with $N=2^{23}$ in $d=2,3$, we measured the winding angles of segments of sizes $128,256,...,2^{22}$, starting from the origin (i.e, the size of the $i$ segment was $64*2^i$). The change in the winding angle $\Delta\theta_i$ and the correlation between the changes of different segments were measured. In both $d=2$ and $d=3$, we find that the correlation between the different segments is very weak (Pearson correlation smaller than 0.05). Thus, to a good approximation, $\Delta\theta_i$ can be considered as independent of each other. In Fig.~\ref{fig:gaussian} we present the variances of the individual segments. The first segment was omitted from the graph since it scales differently than the others (The first segment starts at the origin and does not have a preceding segment that is half its size.). As expected, in $d=2$, the variance of the different segments is constant. We find that it equals 0.75 as denoted by the dashed line in Fig.~\ref{fig:gaussian}. In $d=3$, we see that the variance of the winding angle increases as $i$ increases, as we predicted earlier. The solid line in Fig.~\ref{fig:gaussian} denotes a power law growth of $\langle\Delta\theta_i^2\rangle\sim i^{0.52}$, which is consistent with the results of Section \ref{sec:SAW3d}. (Due to the small range of $i$ and some arbitrariness in the numbering of the segments, the error in this exponent can be as large as 0.1.) Under the assumption that the winding angles of the different segments are independent, 
\begin{equation} \label{eq:GaussSum}
	\langle\theta^2\rangle\approx\int_1^m\langle\Delta\theta_i^2\rangle di \propto m^{1.52}\propto(\ln N)^{1.52}.
\end{equation}

\section{Summary and discussion}
Using a recent implementation of the pivot algorithm,\cite{Clisby2010} we were able to study the winding angle $\theta$ of RWs and SAWs on square and cubic lattices of sizes that were not previously available to simulations. The method described in Sec.~\ref{sec:algo} to compute the winding angle relies on the fact that some properties of a lattice walk can be deduced from aggregate information about large sections that constitute the walk, without knowing its small scale details. This approach can be useful in various situations. For example, it is possible to perform fast intersection tests of a SAW with various surfaces.\cite{Hammer2015, Clisby2016} This can be used in future studies to measure the distribution $p(\theta)$ of the winding angle of long walks near excluded regions of different shapes and sizes. Specifically, it would be interesting to know how a smaller radius of the excluded center increases the winding angle of a SAW. 

By studying RWs and  two-dimensional SAWs with the number of sites $N$ ranging up to $10^7$, we were able to observe the $N$ dependence of $p(\theta)$ that was predicted by the theory. For RWs, we showed that as $N\rightarrow\infty$, $\sqrt{\langle\theta^2\rangle}$ is linear in $\ln N$ with the predicted slope $1/\sqrt{12}$, apart from finite size corrections (Eqs.~\eqref{eq:RW2dNonLinFit} and \eqref{eq:RW2dNonLinFit2}). For SAWs on the square lattice, we showed that the ratio $\langle\theta^4\rangle/\langle\theta^2\rangle^2$ approaches Gaussian value 3, as is predicted by the theory, with a small correction that decays slowly as $N$ increases. 

For SAWs on the cubic lattice, we observed non-monotonic dependence of $\langle\theta^4\rangle/\langle\theta^2\rangle^2$ on $N$. This surprising result shines a different light on the previous result by Walter \textit{et al.},\cite{Walter2011} where it was shown that for SAWs with $N\le25000$, $\langle\theta^4\rangle/\langle\theta^2\rangle^2$ converges to a constant value of 3.74(5) as $N$ is increased. (We show that this is in fact approximately the maximum value of  $\langle\theta^4\rangle/\langle\theta^2\rangle^2$.) This behavior might indicate a cross-over from a non-asymptotic regime to the asymptotic behavior in the limit $N\rightarrow\infty$. It is possible that the cross-over is related to the structure of the lattice. We showed that the lattice structure is evident in the winding angle distribution even for walks with $N=10^5$ and diminishes for larger walks. 

In Sec.~\ref{sec:gussian} we demonstrated that the square winding angle of a SAW in $d=3$ can be obtained from the summation of the square change in the winding angles of $m\propto\ln N$ independent segments of the walk. Unlike the situation in $d=2$, where these segments have identical mean square winding angles, in $d=3$ the mean square winding angle of the $i$ segment increases approximately as $i^{0.52}$, which leads to an increase of the total square winding angle proportional to $(\ln N)^{1.52}$, as was measured here and in Ref.~\onlinecite{Walter2011}. We stipulate that the increase in the winding angle of the individual segments can be explained by the fact that when the segment are scaled down to the same size, the excluded center is also effectively scaled down, and thus the winding angle is increased.

\begin{acknowledgments}
We thank T. A. Witten for enlightening discussions of the subject, and M. Kardar for numerous suggestions during the entire work and for comments on the manuscript. This work was supported by the Israel Science Foundation grant 186/13.
\end{acknowledgments}

\bibliographystyle{apsrev4-1}
\bibliography{windAngRefs}
\end{document}